\documentclass[twocolumn,prl,showpacs,nofootinbib,amsmath,amssymb]{revtex4}

\usepackage{graphicx}
\usepackage{dcolumn}
\usepackage{bm}
\usepackage{color}

\newcommand{\delt}{{\vec\nabla_{\vec\theta}}}

\def\cleb#1#2#3#4#5#6{{\cal C}^{#1#2}_{#3 #4 \,\, #5 #6}}

\def\ALMm{{A^{\ominus}}^{LM}_{ll'}}

\begin{document}

\title{Lensing of 21-cm Fluctuations by Primordial Gravitational
Waves}

\author{Laura Book$^1$, Marc Kamionkowski$^{1,2}$, and Fabian Schmidt$^{1}$}
\affiliation{$^1$California Institute of Technology, Mail Code 350-17,
     Pasadena, CA 91125}
\affiliation{$^2$Department of Physics and Astronomy, Johns
     Hopkins University, 3400 N.\ Charles St., Baltimore, MD 21210}

\date{\today}

\begin{abstract}
Weak-gravitational-lensing distortions to the intensity pattern
of 21-cm radiation from the dark ages can be decomposed
geometrically into curl and curl-free components.  Lensing by
primordial gravitational waves induces a curl component, while
the contribution from lensing by density fluctuations is strongly suppressed.  
Angular fluctuations in the 21-cm background extend to very
small angular scales, and measurements at different frequencies
probe different shells in redshift space.  There is thus a huge
trove of information with which to reconstruct the curl
component of the lensing field, allowing tensor-to-scalar ratios
conceivably as small as $r\sim 10^{-9}$---far smaller than those
currently accessible---to be probed.
\end{abstract}


\maketitle

One of the principle aims of early-Universe cosmology is
detection of the inflationary gravitational-wave (IGW) background
\cite{Rubakov:1982df} via measurement of the curl pattern
\cite{Kamionkowski:1996ks}
that it induces in the cosmic microwave background (CMB) polarization.
Likewise, a principle aim of physical cosmology is measurement
of the distribution of atomic hydrogen during the ``dark ages,''
the epoch after recombination and before the formation of the first stars and
galaxies, via detection of hydrogen's 21-cm line
\cite{Loeb:2003ya,Furlanetto:2006jb,arXiv:0910.3010}.  Several
experiments are poised to soon detect the 21-cm signal from the
epoch of reionization \cite{lofar}, and there are longer-term
prospects to delve into the dark ages  \cite{arXiv:0902.0493}.
In this paper, we show that angular fluctuations of the 21-cm
intensity may ultimately provide an IGW probe that extends to
amplitudes smaller than those currently accessible with the CMB. 

Weak gravitational lensing of galaxies by large-scale density
perturbations
\cite{Blandford:1991zz} was
detected in 2000 \cite{Bacon:2000sy} and is now a chief aim of a
number of ongoing and future galaxy surveys.  These efforts seek
the lensing-induced distortions
of galaxy shapes.  Weak lensing of the CMB by density
perturbations was detected recently \cite{Smith:2007rg}.
The observational signatures here are lensing-induced
position-dependent departures from statistical isotropy in the
two-point CMB correlation functions, or equivalently, the
four-point correlation functions induced by lensing
\cite{Seljak:1998aq}.

Primordial gravitational waves can likewise lens both galaxies
and the CMB
\cite{Kaiser:1996wk,Cooray:2005hm,Book:2011na}.
The most general
lensing pattern can, like the CMB polarization, be decomposed
into curl and curl-free parts \cite{Kamionkowski:1997mp}.  Since
density perturbations produce (to linear order in the deflection
angle) no curl in the lensing pattern, the curl component
provides an IGW probe.  The problem,
however, is that the curl signal, even with the most optimistic
assumptions about IGWs, is well
below the noise for both current galaxy surveys and even for
optimistic next-generation CMB experiments.

Here we consider lensing of intensity fluctuations in the 21-cm
signal from atomic hydrogen in the dark ages.
Atomic hydrogen in the redshift range $30\lesssim z \lesssim
200$ can absorb radiation deep in the Rayleigh-Jeans region of
the CMB \cite{Loeb:2003ya}.  Measurement of this absorption,
over some narrow frequency range (corresponding to a narrow
redshift range), over the sky thus maps the spatial distribution
of hydrogen at that redshift.  The angular power spectrum of
these 21-cm fluctuations extends to multipole moments $l \sim
10^7$ (limited only by the baryonic Jeans mass)
\cite{Loeb:2003ya}, far larger than those, $l\sim3000$, to which
the CMB power spectrum extends (beyond which fluctuations are
suppressed by Silk damping).  
The signatures of gravitational lensing of
these 21-cm angular correlations are precisely the same as those
of lensing of the CMB temperature map---local departures from
statistical isotropy.  We can therefore adopt unchanged the
mathematical formalism for lensing of the CMB.

Our work resembles in spirit that in Ref.~\cite{Masui:2010cz}
which argued that the huge number of Fourier modes available in
21-cm maps of the dark-age hydrogen distribution would provide considerable 
statistical significance in detecting the IGW
distortion to matter fluctuations.  However, they consider 
the {\it intrinsic} distortion to matter fluctuations by IGWs.
On the other hand, we consider the
distortion to the {\it images} of the matter distribution by
lensing by IGWs.  Our work is related to that of
Ref.~\cite{Sigurdson:2005cp}, who considered reconstruction of
the lensing field due to density perturbations with 21-cm
fluctuations.

\begin{figure}[htbp]
\includegraphics[width=3.4in]{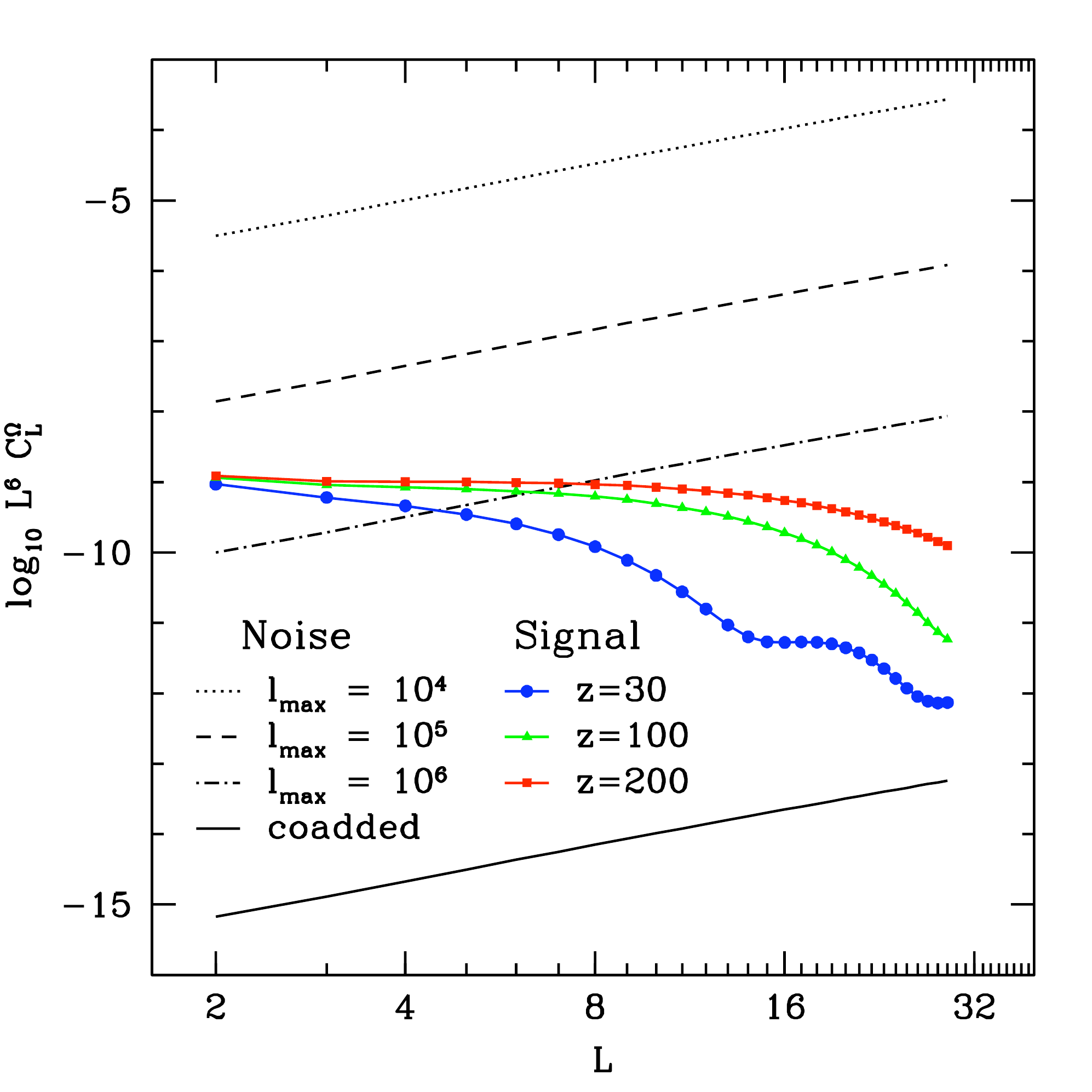}
\caption{The power spectrum for the deflection-field curl
     component for lensing of sources at various redshifts by a
     scale-invariant spectrum of IGWs of the largest amplitude ($r=0.2$)
     consistent with current measurements.  We also superimpose
     noise power spectra for lensing reconstruction carried
     out to various values of $l_{\rm max}$.  Also shown is the
     noise power spectrum we estimate from co-adding the signals
     from all possible redshifts, assuming an $l_{\rm max} = 10^6$.}
\label{fig:CLOmega}
\end{figure}

The most general deflection field $\vec{\Delta}$ can be
written as a function of position $\hat n$ on the sky as
\cite{Kamionkowski:1997mp},
\begin{equation}
     \vec{\Delta} = \delt \, \phi(\hat{n}) + \delt \times
     \Omega(\hat{n}),
\label{eqn:displacements}
\end{equation}
in terms of curl-free ($\delt \phi$) and curl ($\delt\times
\vec{\Omega}$) components.  The angular power spectrum for the curl
field $\Omega(\hat n)$ due to lensing of sources at redshift $z$
by IGWs with power spectrum $P_T(k)$ is
\begin{equation}
     C_L^{\Omega} = 2\int \, \frac{d^3k}{(2\pi)^3} P_T(k)
     \left[F_L^X(k)\right]^2,
\label{eqn:phiOmegapowerspectra}
\end{equation}
where
\begin{equation}
     F_L^{\Omega}(k) = - \sqrt{ \frac{2 \pi (L+2)!}{(L-2)!}}
     \int_{k\eta(z)}^{k\eta_0} \frac{T(w)}{L(L+1)} \frac{
     j_L(k\eta_0-w)} {(k\eta_0-w)^2} dw, 
\label{eqn:Omegatransfer}
\end{equation}
and $\eta_0$ and $\eta(z)$ are the conformal time today and at
redshift $z$, respectively.  Here $T(w)\simeq 3 j_1(w)/w$ is
the gravitational-wave transfer function, and $j_n(x)$ are the
spherical Bessel functions.  The angular power spectra for
the lensing of sources at several redshifts are shown in
Fig.~\ref{fig:CLOmega}; for $L\lesssim 6$, the source-redshift 
dependence is weak for a scale-invariant gravitational-wave background.

We now review how this power spectrum is measured
following the treatment of lensing of the CMB in
Ref.~\cite{Book:2011na}, focusing on a single redshift slice first.  
Given a map $I(\hat n)$ of the 21-cm intensity as a function of position
$\hat n$ on the sky, the minimum-variance estimator for the
spherical-harmonic coefficients for the curl component of
lensing is
\begin{equation}
     \widehat{\Omega_{LM}} = \frac{\sum_{ll'} Q^{\ominus
     L*}_{ll'} \widehat{\ALMm}\big/\left( 
     C^{\rm map}_l C^{\rm map}_{l'} \right)}{\sum_{ll'}
     \left|Q^{\ominus L}_{ll'}\right|^2/\left( C^{\rm map}_l
     C^{\rm map}_{l'} \right)},
\label{eqn:estimator}
\end{equation}
where $C_l^{\rm map}=C_l+C_l^{\rm n}$ is the angular power spectrum of
the map with $C_l$ the power spectrum of the 21-cm
intensity and $C_l^{\rm n}$ the noise power
spectrum, and the sums are only over $l+l'+L=$odd.
We use lower-case $l$ for CMB fluctuations
and upper-case $L$ for the lensing-deflection field.  Here,
\begin{align}
     Q^{\ominus L}_{ll'} =\:& \frac{i}{\sqrt{2L+1}} \left[
     \frac{C_l G^L_{l'l}}{\sqrt{l'(l'+1)}} - \frac{C_{l'}
     G^L_{ll'}}{\sqrt{l(l+1)}}\right], \nonumber\\
     G_{ll'}^L \equiv\:& \sqrt{\frac{L(L+1)l(l+1)l'(l'+1)
     (2l+1)(2l'+1)}{4\pi}} \cleb{L}{1}{l}{0}{l'}{1},
     \nonumber\\
     \widehat{\ALMm} &= \sum_{mm'} a^{\rm
     map}_{lm} a^{*\,{\rm map}}_{l'm'} (-1)^{m'}
     \cleb{L}{M}{l}{m}{l',}{-m'},
\label{eqn:biposh}
\end{align}
where $\widehat{\ALMm}$ 
are estimators for odd-parity bipolar-spherical-harmonic coefficients
\cite{Hajian:2003qq} in terms of the
spherical-harmonic coefficients $a_{lm}^{\rm map}$ of the 21-cm
map and Clebsch-Gordan coefficients
$\cleb{L}{M}{l}{m}{l',}{-m'}$.  The estimator for the power
spectrum of the curl component of the deflection
field is then $\widehat{C_L^{\Omega}} = \sum_m
|\widehat{\Omega_{LM}}|^2 / (2L+1)$.  The variance of
$\widehat{\Omega_{LM}}$  under the null hypothesis is given by
\begin{equation}
     \left(\sigma^{\Omega}_L\right)^2 \equiv \left\langle
     |\widehat{\Omega_{LM}}|^2 \right\rangle = 2
     \left[\sum_{ll'}\left| Q^{L\ominus}_{ll'}\right|^2/\left(
     C^{\rm map}_l C^{\rm map}_{l'}
     \right)\right]^{-1}.
\label{eq:Omvar}
\end{equation}
This noise power spectrum is plotted in Fig.~\ref{fig:CLOmega} using the
21-cm power spectra from Ref.~\cite{Loeb:2003ya} and taking
the noise power spectrum $C_l^{\rm n}=0$ for $l<l_{\rm max}$ and
$C_l^{\rm n}=\infty$ for $l>l_{\rm max}$.  We show results for several $l_{\rm
max}$ which are, roughly speaking, the maximum value of $l$ with
which the 21-cm power spectrum can be measured with high
signal-to-noise.  The signal-to-noise (squared) with which IGWs
can be detected is then
\begin{equation}
      (S/N)^2 =
      \sum_L \, (L+1/2) \left(C_{L}^{\,\Omega}
      \right)^2 / (\sigma_L^\Omega)^4.
\label{eqn:SN}
\end{equation}

Before reviewing the numerical results, it is instructive to
consider an analytic estimate of the noise power spectrum
$\left(\sigma_L^\Omega \right)^2$.  To do so, we use the flat-sky
approximation \cite{Cooray:2005hm},
\begin{equation}
     \left(\sigma_L^\Omega \right)^{-2} = \int \frac{d^2 l}{(2\pi)^2}
     \frac{ (\vec L \times \vec l)^2 (C_l - C_{|\vec L-\vec
     l |})^2}{ 2 C_l^{\rm map} C_{|\vec L-\vec l|}^{\rm map} }.
\label{eqn:flatsky}
\end{equation}
For $L \ll l$ we approximate $|\vec L-\vec l| \simeq l-L\cos\alpha$,
where $\cos\alpha\equiv \hat L\cdot \hat l$, and
$C_{|\vec L-\vec l|} \simeq C_l - L (\cos\alpha) (\partial
C_l/\partial l)$.  If $C_l\propto l^n$, then
\begin{eqnarray}
     \left(\sigma_L^\Omega \right)^{-2} &=& \int \frac{l\, dl}{4\pi^2}
     \int_0^{2\pi} d\alpha\frac{1}{2} L^4
     \sin^2\alpha\cos^2\alpha \left(\frac{\partial\ln
     C_l}{\partial \ln l}\right)^2
     \nonumber \\
     & \simeq &  L^4 n^2 l_{\rm max}^2 /(64 \pi).
\label{eqn:sigmaeqn}
\end{eqnarray}
The flat-sky calculation is accurate for $L\gtrsim 20$ and 
\emph{over}estimates the noise by up to 30\% at smaller $L$.  
As shown in Fig.~2 in Ref.~\cite{Loeb:2003ya}, the 21-cm power
spectrum extends without suppression out to $l\gtrsim 10^6$, and
values of $l_{\rm max}\sim10^7$ are perhaps achievable with a bit
more effort. However, given the rapid suppression
of the 21-cm power spectrum at higher $l$, the return on the
investment of noise reduction in terms of higher $l_{\rm max}$
will probably be small above $l_{\rm max} \simeq 10^7$.

We now approximate the $\Omega$ power spectrum (for $r=0.2$) as
$C_L^{\Omega} \simeq 10^{-11}\, (L/2)^{-6}$.  Although
this approximation differs from the numerical results for
different redshifts $z$ at $L\simeq 30$, it is quite accurate
for all $30\lesssim z \lesssim200$ for the smallest $L$ where most 
of the signal arises.  From
Eq.~(\ref{eqn:SN}), the signal-to-noise with which the
gravitational-wave background can be detected is
\begin{equation}
     (S/N) \simeq 4.5\, \left( l_{\rm
     max}/10^6 \right)^2 \left(n/2 \right)^2
     \left(L_{\rm min}/2 \right)^{-1},
\label{eqn:singlez}
\end{equation}
where $L_{\rm min}$ is the minimum $L$ that can be measured.

There are several things to note about this result:
(1) The signal-to-noise obtained with the adopted fiducial
values for $l_{\rm max}$, $L$, and $n$ is significant. (2)
The scaling of the signal-to-noise with $l_{\rm max}$ is very
rapid, and greater than what might have been expected ($\propto
l_{\rm max}$) naively.  The origin of this rapid scaling is
similar to that for detection of the local-model trispectrum
\cite{Kogo:2006kh} (as the signal we are measuring here is,
strictly speaking, an intensity trispectrum).  Thus, the
sensitivity to a gravitational-wave background increases very
rapidly as the angular resolution of the map is improved.  (3)
The sensitivity decreases as $L_{\rm min}$ is increased, so good
sky coverage is important for gravitational-wave detection.

While a signal-to-noise of 4.5 is respectable, and could be
improved with even larger $l_{\rm max}$, we
can go much further:  By changing the frequency at which the
21-cm map is made, we look at spherical shells of atomic
hydrogen at different redshifts.  Suppose, then, that we have
21-cm maps at two different frequencies that correspond to
spherical shells separated along the line of sight by a comoving
distance $\delta R$.  Those two maps are statistically
independent at the highest $l$ (where the vast majority of the
signal-to-noise for IGW detection arises) if
$(\delta R/R) \gtrsim l^{-1}$.  If $\Delta R$ is the separation in
comoving radius corresponding to the entire frequency range covered
by the observations (say, redshifts $z\simeq30 - 200$), then the 
total number of statistically
independent maps that can be obtained is $N_z \simeq
(\Delta R/\delta R) \simeq l(\Delta R/R) \simeq 0.15\,l$.
If so, then the signal-to-noise from all these
redshift ranges can be added in quadrature, and the
signal-to-noise then increases by a factor $N_z^{1/2}$.  But
there may be room for even more improvement:  If
most of the lensing occurs at redshifts $z\lesssim30$ (as is the
case for the lowest $L$), then the lensing pattern is the same
for all redshift shells in which case every redshift shell
contributes coherently to an estimator for $\Omega_{LM}$.  In
this case, $(\sigma_L^\Omega)^2$ is decreased by factor
$N_z^{-1}$, and the signal-to-noise increased by a factor $N_z$
relative to the single-$z$ estimate.  Since
most of the signal comes from the lowest $L$, we estimate that
the signal-to-noise for IGW detection obtained
by coadding redshift shells will be
\begin{equation}
     \left(S/N \right)_{\rm tot} \simeq 6.8\times10^5\, \left( l_{\rm
     max}/10^6 \right)^{3} \left(n/2 \right)^2
     \left( L_{\rm min}/2 \right)^{-1},
\label{eqn:multiz}
\end{equation}
assuming (as above) the largest currently allowed IGW
amplitude $r\simeq 0.2$.
Put another way, the smallest tensor-to-scalar ratio that can be
detected at the $3\sigma$ level is
\begin{equation}
     r\simeq 10^{-6} \left( L_{\rm min}/2 \right)
     \left(l_{\rm max}/10^6 \right)^{-3}
     \left(n/2 \right)^{-2}.
\label{eqn:smallest}
\end{equation}
Note that the dependence on $l_{\rm max}$ is very steep, and
including all the information to $l_{\rm max} = 10^7$ could yield a 
detection threshold of $r \simeq 10^{-9}$.  
The full-sky calculation, including a more realistic shape of $C_l$, yields
a result consistent with this estimate (Fig.~\ref{fig:CLOmega}).  

To put this result in perspective, we note that the current
upper bound $r\lesssim 0.22$ comes from WMAP measurements of
temperature-polarization correlations, although not from B-mode
null searches.  The forthcoming
generation of sub-orbital B-mode experiments are targeting
$r\lesssim 0.1$, and a dedicated CMB-polarization satellite
might then get to $r\sim 10^{-2}$ \cite{Bock:2009xw}.  

Measurement of gravitational-wave amplitudes $r\lesssim 0.01$
with CMB polarization will have to contend with the additional
contribution to B-mode polarization from gravitational lensing
(by density perturbations) of primordial E modes
\cite{Zaldarriaga:1998ar}.  The two
contributions (IGW and lensing) to
B modes can be distinguished if the lensing deflection angle can be
reconstructed with small-scale CMB fluctuations
\cite{Kesden:2002ku,Knox:2002pe}.  This may allow values $r\sim
10^{-3}$ to be probed, although it requires a far more
sophisticated CMB experiment (with far better angular
resolution) than simple detection of B modes would require.

Further progress in separation of lensing and
IGW contributions to B modes can be obtained with
21-cm measurements \cite{Sigurdson:2005cp} of precisely the
type we discuss here but of the curl-free
lensing component (due to density perturbations) rather than the
curl component from IGWs.  Such measurements,
when combined with a precise CMB polarization experiment, can in
principle get to IGW amplitudes comparable to those
we have discussed here.  Measurement of the 21-cm curl component
may therefore ultimately be competitive for the most sensitive
probe of IGWs, even if a sensitive CMB-polarization experiment
is done.  Furthermore, if both 21-cm observations and a
CMB-polarization map are available, then measurement of the
21-cm curl component can be used as a cross-check and to
complement a measurement from the combination of B-mode
polarization with 21-cm lensing subtraction.

While we have focussed here on the dark ages, similar
measurements can also be performed with 21-cm fluctuations from
the epoch of reionization and also with galaxy surveys; the
critical issue will be how high $l_{\rm max}$ can get.
While the 21-cm curl component induced by lensing by
density perturbations at second order is too small to be an
issue \cite{Cooray:2005hm}, a curl component may conceivably
arise since the atomic-hydrogen distribution is not perfectly
Gaussian due to non-linear gravitational collapse and baryonic
effects.  We
speculate that this curl component will be small for the
small-$L$ modes at which the IGW signal peaks.  We also
imagine that the information from multiple redshifts may be
combined to separate the IGW and any bias-induced
signal.

To close, we note that the measurements we describe will be
challenging and are very futuristic compared to what current and
next-generation experiments will accomplish.  Still,
21-cm cosmology is an exciting and rapidly developing
experimental arena, for a good number of scientific reasons
\cite{Furlanetto:2006jb}, and we hope that the idea presented
here provides one additional motivation to carry such work
forward.

\bigskip

This work was supported by DoE DE-FG03-92-ER40701, NASA
NNX10AD04G, and the Betty and Gordon Moore Foundation.

\end{document}